\documentclass{aav6_nc}

\usepackage{txfonts}
\usepackage{graphicx}
\usepackage{natbib}

\begin{document}

\title{Searching for p-modes in MOST\thanks{Based on data from the MOST
satellite, a
Canadian Space Agency mission, jointly operated by Dynacon Inc.,
the University of Toronto Institute for Aerospace Studies and
the University of British Columbia, with the assistance of
the University of Vienna.} Procyon data: Another view}

\author{F. Baudin\inst{1}
\and T. Appourchaux\inst{1} 
\and P. Boumier\inst{1}
\and R. Kuschnig\inst{2}
\and J.W. Leibacher\inst{1,3}
\and J.M. Matthews\inst{2}}

\offprints{F. Baudin, {\tt {\small frederic.baudin@}}
{\tt {\small ias.u-psud.fr}}}

\authorrunning{Baudin {\it et al.}}
\titlerunning{Another view of MOST Procyon data}

\institute{Institut d'Astrophysique Spatiale, CNRS/Universit\'e Paris XI
UMR 8617, F-91405 Orsay, France
\and Department of Physics and Astronomy, University of British Columbia,
6224 Agricultural Road, Vancouver V6T 1Z1, Canada
\and National Solar Observatory, 950 N. Cherry Avenue,
Tucson AZ 85719-4933, USA}

\date{Received ; Accepted}

\abstract
{Photometry of Procyon obtained by the MOST satellite
in 2004 has been searched for p~modes by several groups, with
sometimes contradictory interpretations.}
{We explore two possible factors that complicate the analysis and
may lead to erroneous reports of p~modes in these data.}
{Two methods are used to illustrate the role of subtle
instrumental effects in the photometry: time-frequency analysis,
and a search for regularly spaced peaks in a Fourier spectrum
based on the echelle diagramme approach.}
{We find no convincing evidence of a p-mode signal in the
MOST Procyon data. We can account for an apparent excess of power
close to the p-mode frequency range and signs of structure in an
echelle diagramme in terms of 
instrumental effects.}
{}

\keywords{Stars: Procyon -- Stars: oscillations}  

\maketitle

\section{Introduction}

The search for acoustic oscillations in the star Procyon A
(which we call hereafter simply Procyon) has been a key
part of the young field of stellar seismology for two decades,
since the first claim of p-mode detection by \citet{bernard86}.
In recent years, interest grew rapidly with the announcement by
\citet{martic99} of global acoustic modes in Procyon detected
in high-resolution, time-resolved spectroscopy, confirmed by
\citet{caro}.  Interest turned to controversy with the
non-detection of intensity oscillations by the MOST
(Microvariability \& Oscillations of STars) space-borne
photometer \citep{jaymie}, while contemporaneous
groundbased spectroscopic campaigns by \citet{martic04},
\citet{eggen04} and \citet{bouchy04} appeared to
confirm the presence of stellar oscillations. This non-detection
triggered considerable speculation about non-stellar noise
sources in the MOST data \citep{jcd04,bedding05}
and independent analyses of those data
claiming possible evidence for p~modes, but using different
analysis methods \citep{clara,garcia05}.
The contrast between the reported p-mode signal in luminosity
and integrated radial velocity also inspired theorists to
refine models of convection and turbulence in Procyon
\citep[e.g.,][]{robinson2005}.

The MOST team has described the nature of the Fabry Imaging data
on which the Procyon photometry is based \citep{reegen2006},
including the recognised effects of stray light due to
Earthshine modulated by the MOST satellite orbit and a
beating effect of the science and startracker CCD timing in the
early stages of the mission.

We present an independent analysis
of the MOST Procyon data compared to simulations, calling
attention to how the timing effects and orbital artifacts can
be misinterpreted as evidence for p-mode oscillations.

\begin{figure*}
\centering
\hbox{\includegraphics[height=9cm,angle=90]{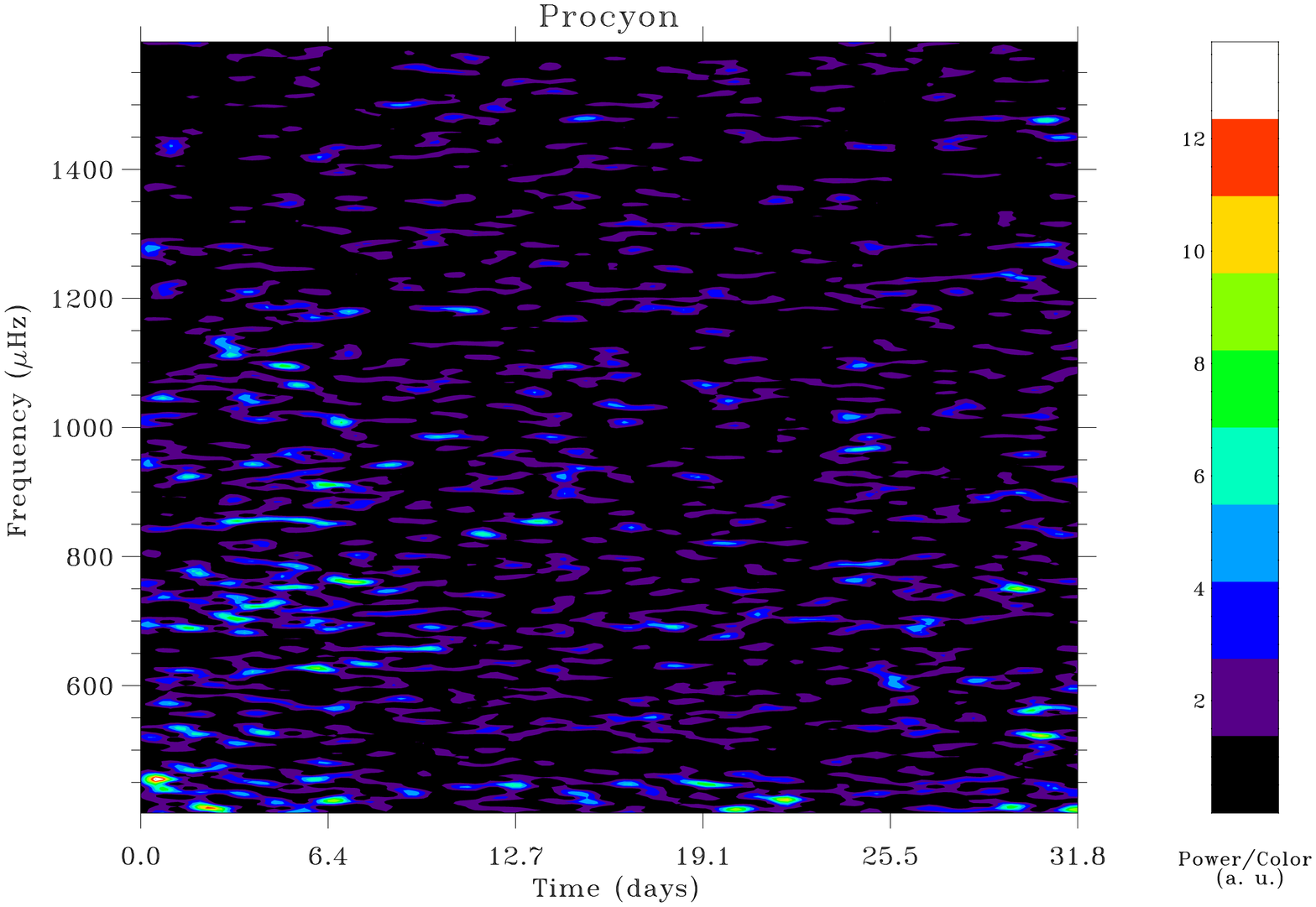}
\includegraphics[height=9cm,angle=90]{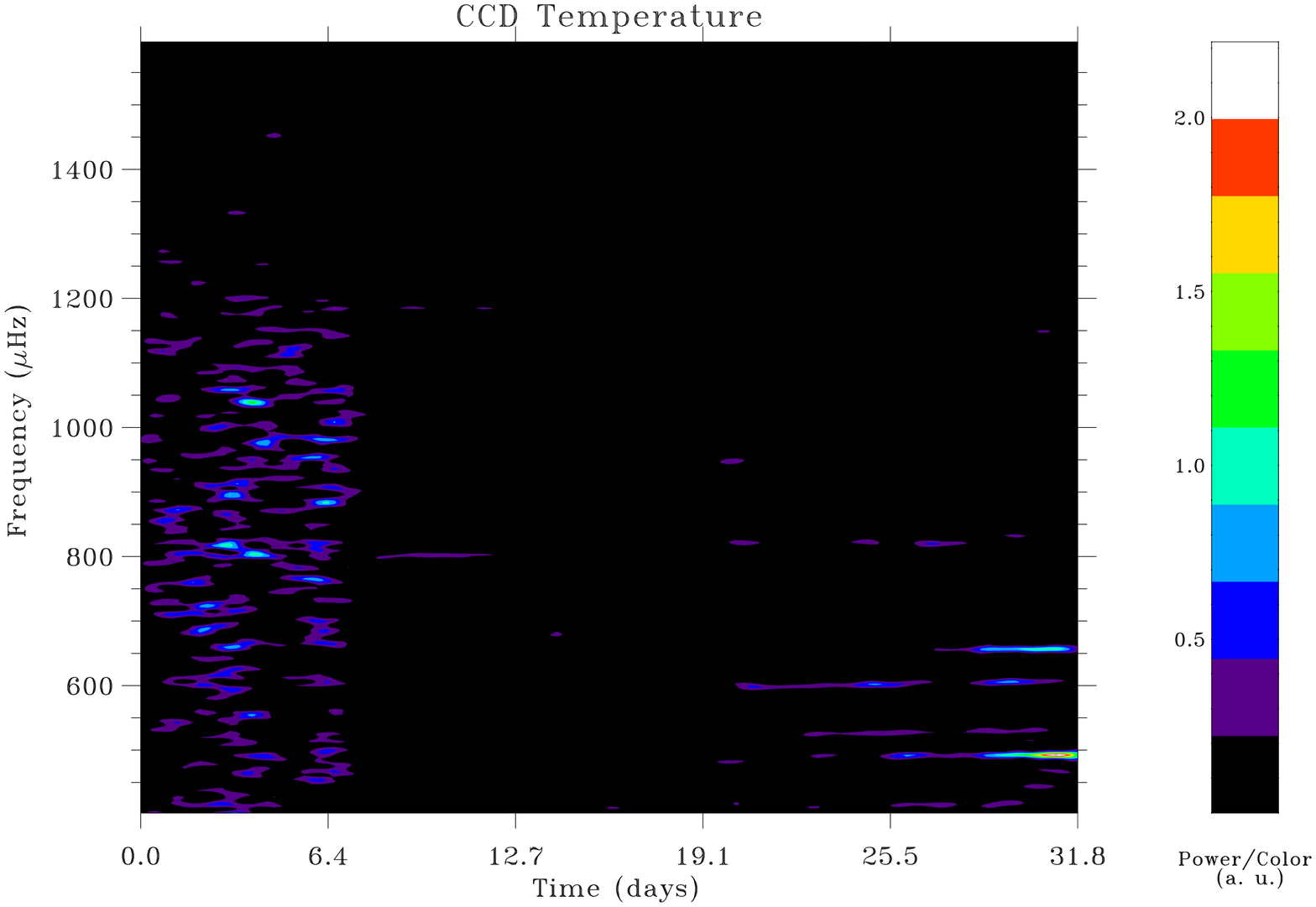}}
\caption{Left: Time/frequency diagramme of Procyon data, showing excess
power between 500 and  1200~$\mu$Hz during the first eight
days of observation.
Right: same analysis of the CCD temperature, showing the same
excess of power}
\label{fig:tempsfreq}
\end{figure*}

\section{Time-frequency analysis}

\begin{figure}
\centering
\includegraphics[height=8.5cm,angle=90]{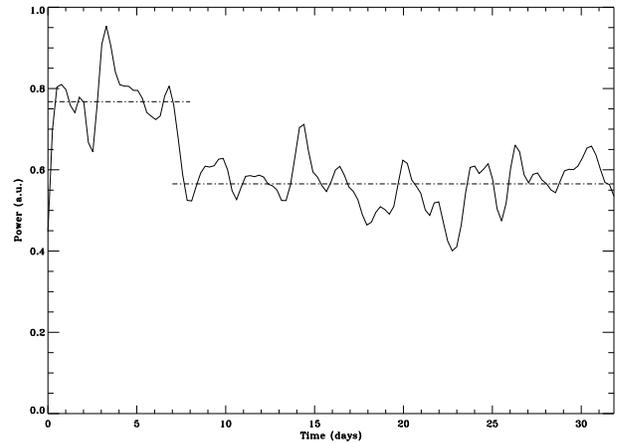}
\caption{Time variation of the integrated power (from 400 to 1600~$\mu$Hz),
showing two different levels during the observations.}
\label{fig:John}
\end{figure}

\begin{figure}
\centering
\includegraphics[height=8.5cm,angle=90]{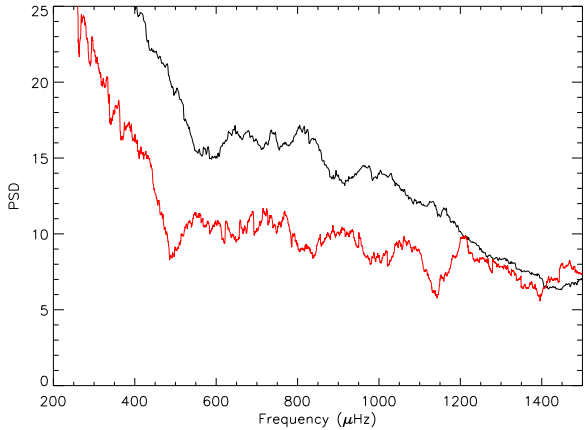}
\caption{Smoothed Fourier spectra of Procyon data for the first seven
days (black) and last 25 days (red) of observation}
\label{fig:procyon_tf}
\end{figure}

The 32-day-long MOST photometric time series has a duty cycle
of 99\% \citep[for more details, see][]{jaymie}. This
continuity of the data, currently unique to MOST observations
of stars and a key element of the motivation for future
photometric space missions such as CoRoT and Kepler, has many
important advantages.  It eliminates the cycle/day aliases
common to groundbased campaigns, even with multiple sites.
Another benefit, which we exploit in this paper, is that these
data make it possible to investigate variation of the frequency
content of the signal as a function of time more sensitively
than for other astronomical data sets.\\
We have applied a time-frequency analysis, based on a Morlet
wavelet \citep[see][]{baudin94}, to the time series of the
MOST photometry and to time series of instrumental parameters
such as satellite pointing error and CCD temperature, which
are included in the MOST science telemetry.
The left-hand panel of Figure~\ref{fig:tempsfreq} shows the distribution of
power in the MOST Procyon photometry as a function of both
frequency and time.  There is some excess power during the
first 8 days of the observations, in the frequency range of
about 500\,--\,1200 $\mu$Hz.  The contrast in excess power is
small and can be seen more clearly when the power is summed
over frequency as a function of time, as shown in Figure~\ref{fig:John}.
The average increase in the excess power is about 35\%.
The same time-frequency analysis was applied to time series
of spacecraft and instrument parameters in the MOST science
telemetry. Some (but not all) temperatures measured on board
seem to show an excess of power in the same ranges of time
and frequency as in the stellar light curve.  The
time-frequency plot of the MOST science CCD temperature is shown
in the right-hand panel of Figure~\ref{fig:tempsfreq}.  Other
parameters, such
as telescope--pointing wander, do not show such behaviour.
However, this does not mean that on board temperature
variations are the cause of the observed photometric power
excess. MOST CCD temperatures (operating at around {\mbox -35$^o$C})
are controlled to within 0.1$^o$C and are monitored to an
accuracy of 0.01$^o$C. The very small amplitude of the
temperature variations is insufficient to change the dark
current or other characteristics of the CCD output to account
for the excess seen in Figure~\ref{fig:John}.  Furthermore, there is no
one-to-one correlation in time and frequency.  More likely,
both the temperature and photometric variations are symptoms
of a common cause.  There is nothing in the MOST operations
records or in other telemetry that can explain such a
change on the seventh day of the observations. Among the different
instrumental effects described by \cite{reegen2006}, none seems to be clearly
the cause of this observation.
However, one must note
that these seven days represent the first seven days of science
operations of the MOST satellite.\\
The smoothed Fourier spectrum of the first seven days of the
Procyon photometry shows greater spectral power density up
to about 1200 $\mu$Hz than during the remaining 24 days (Figure~\ref{fig:procyon_tf}).
This includes the frequency range where acoustic vibrations
of the star are anticipated, both from theory and the
spectroscopic observations. Therefore, this power contrast
must be taken into account in any analysis looking for p~modes,
which should be sought preferentially in the second part of the
photometric time series.

\section{Collapsed echelle diagramme analysis}

The acoustic spectrum of a star such as Procyon is expected
to have modes with a roughly regular spacing in frequency
(called the ``large spacing'' ${\Delta}_0$\,). One way to see
this spacing (as well as fine spacings between modes) is
to construct an echelle diagramme \citep{grec}, widely used
in helio- and asteroseismology.  This is done by cutting
the Fourier spectrum into $N_c$ frequency intervals of
width $\Delta$, and then plotting each interval above the
previous one, such that the $x$-axis is the frequency
modulo $\Delta$ and the $y$-axis is the frequency. The
regular spacing of modes in the aymptotic regime yields
clear, nearly vertical, ridges in the echelle diagramme.
To enhance the signal-to-noise ratio (SNR) in an echelle
diagramme, it is possible to collapse the second dimension
(the order of the echelle) as proposed by \citet{skorz}.
If the spacing $\Delta$ chosen
is the actual large spacing ${\Delta}_0$ of the p-mode
eigenspectrum of the star, then a peak should appear or become
clearer in this ``collapsed echelle diagramme''.  To
search for a p-mode spacing in data of low SNR, this can
be repeated for a set of $\Delta$ values and a 2D collapsed
echelle diagramme can be constructed. The resulting
diagramme, built with prewhitened spectra (by dividing
the original spectra by the smoothed --200~$\mu$Hz window-- spectra),
is normalised as follows:
\begin{equation}
	P(\nu)=(\frac{p(\nu)}{2 N_c S_f}-1)/s \label{eq_proba}
\end{equation}
where $s$ is a detection level defined below,
$P(\nu)$ is the normalized, collapsed echelle diagramme; $p(\nu)$ is the
original collapsed echelle diagramme having a $\chi^{2}$ with 2\,$N_{c}
S_{f}$ degrees of freedom, $N_c$ being the number of frequency intervals
collapsed; and $S_f$ is the smoothing factor used, if any. The
normalisation factor ($2 N_c S_f$) in Equation~\ref{eq_proba} allows one to
compensate for the varying number of degrees of freedom in
the collapsed echelle diagramme.\\
We first set the {\it a priori} detection probability such that
we can derive the detection level $s$ using the cumulative
probability for the statistics mentionned above. The
detection level $s$ \citep[ Equation~1]{appour04} is then
given by:
\begin{equation}
	{\cal P}(s' \geq s, q)=\int_{s}^{+\infty} \frac{1}{\Gamma(q)}
	\frac{u^{q-1}}{S^{q}}e^{-u/S}{\rm d}u
\end{equation}
with $q=N_c S_f$ and $S$ is the mean noise level in the power spectrum.
The detection level (\,$s$\,) is set {\it a priori} to be 1\%,
{\it i.e.} there is a one percent chance that in a window of
width $N_w$ bins, a peak due to noise will be higher than the
level $s$. When there is no smoothing factor, the number
of independent bins is $N_{w}$. When a smoothing factor is
applied, the number of independent bins decreases linearly
with the smoothing factor. The level $s$ for a single bin is
then obtained by solving the following equation:
\begin{equation}
	{\cal P}(s' \geq s, q)=0.01\frac{S_f}{N_{w}}
\end{equation}
This approximation is valid because the probability of 1\% is small
compared to one.\\ It should be noted that if a spacing $\Delta_0$
exists in the spectrum, a different spacing $\Delta^\prime_0$ defined as
\begin{equation}
m \Delta^\prime_0 = n\Delta_0 ,
\label{eq_spacing}
\end{equation}
with $m$ and $n$ being integers, will also be observed in an echelle
diagramme. This artifact plagues any technique based on the detection
of regularly-spaced peaks.\\
We apply this method to simulated data in the next
subsection, in the framework of preparation for the
CoRoT mission \citep[see][]{ta_corotbook} in order to detect low-SNR modes with a
regular frequency spacing. We then apply it to the MOST
Procyon photometry.

\subsection{Simulated data}

\begin{figure}
\centering
\includegraphics[height=9cm,angle=90]{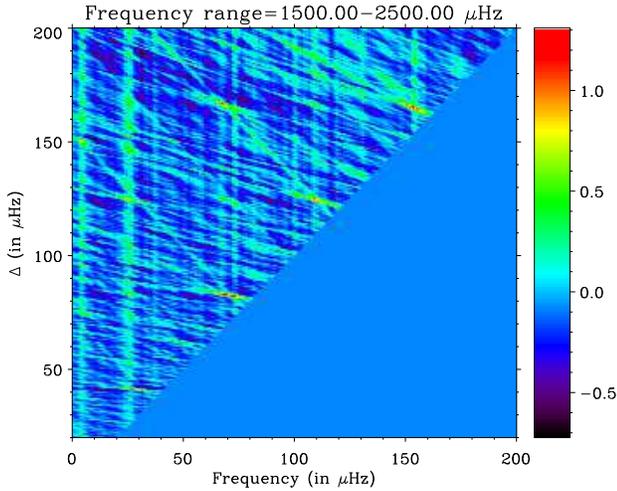}
\caption{Collapsed echelle diagramme for simulated data, showing the signature of
regular spacing of $\simeq$\,82\,$\mu$Hz and multiples of this spacing as defined
in Equation~\ref{eq_spacing}}
\label{fig:hh4}
\end{figure}

The simulated data used here \citep[see][]{fb_corotbook} include
an estimate of the
granulation noise (the dominant component of the intrinsic
stellar noise in the p-mode frequency range) and
theoretically predicted p-mode amplitudes. The granulation
noise is modelled in a simple way (a Lorentzian shape in the Fourier
spectrum) and the modes' amplitudes are based on \cite{samadi03}.
These amplitudes can be attenuated or enhanced
in order to produce a desired SNR.
In the simulations
here, we set the mode SNR (defined as ($P_{\rm mode} +
P_{\rm noise}) / P_{\rm noise}$) to be 1.3\,. The input
frequencies of the simulated modes correspond to a 
mean							
large spacing of ${\Delta}_0 = 82 \mu$Hz,
with a standard deviation of $\simeq 2.3\mu Hz$. A smoothing 	
factor corresponding to $\simeq 3\mu Hz$ has been used.		
Figure~\ref{fig:hh4} shows the collapsed echelle diagramme for this
simulation in which clear peaks above the normalised value
of 1.0 (indicating a 99\% confidence level) appear for
regularly-spaced frequencies having $\Delta \simeq$ 41, 82,
123, and 165 $\mu$Hz.  These correspond to the input value
of the large spacing ${\Delta}_0$ and the corresponding
values ${\Delta}'_0$ (defined in Equation~\ref{eq_spacing}), demonstrating
the strong potential of this approach.

\subsection{MOST Procyon data}

\begin{figure}
\centering
\includegraphics[height=9cm,angle=90]{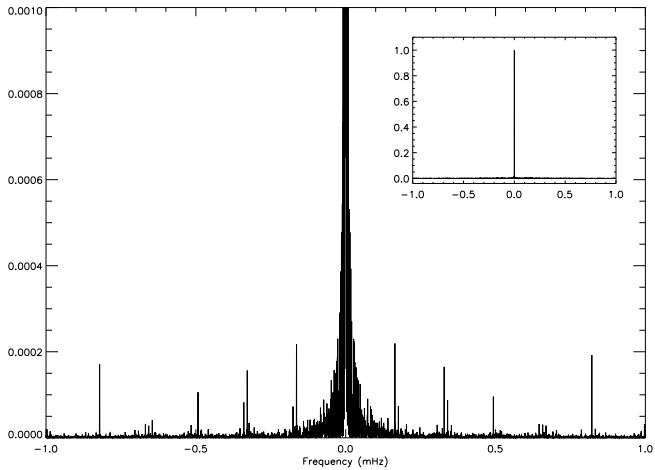}
\caption{Power spectrum of the window function of the observation, almost perfect
but presenting some weak but regularly-spaced peaks}
\label{fig:win}
\end{figure}

\begin{figure*}
\centering
\hbox{\includegraphics[height=8.5cm,angle=90]{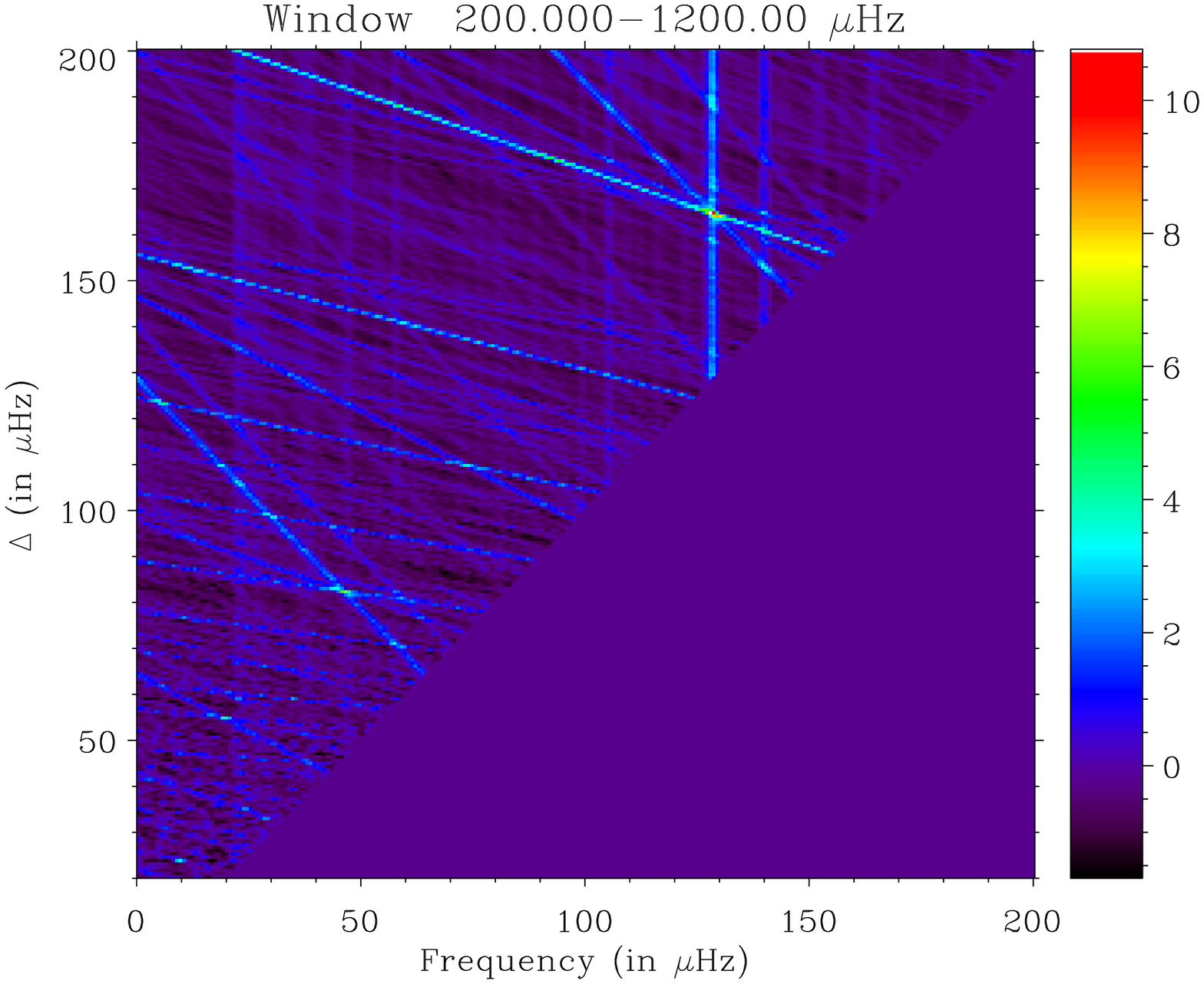}
\includegraphics[height=8.5cm,angle=90]{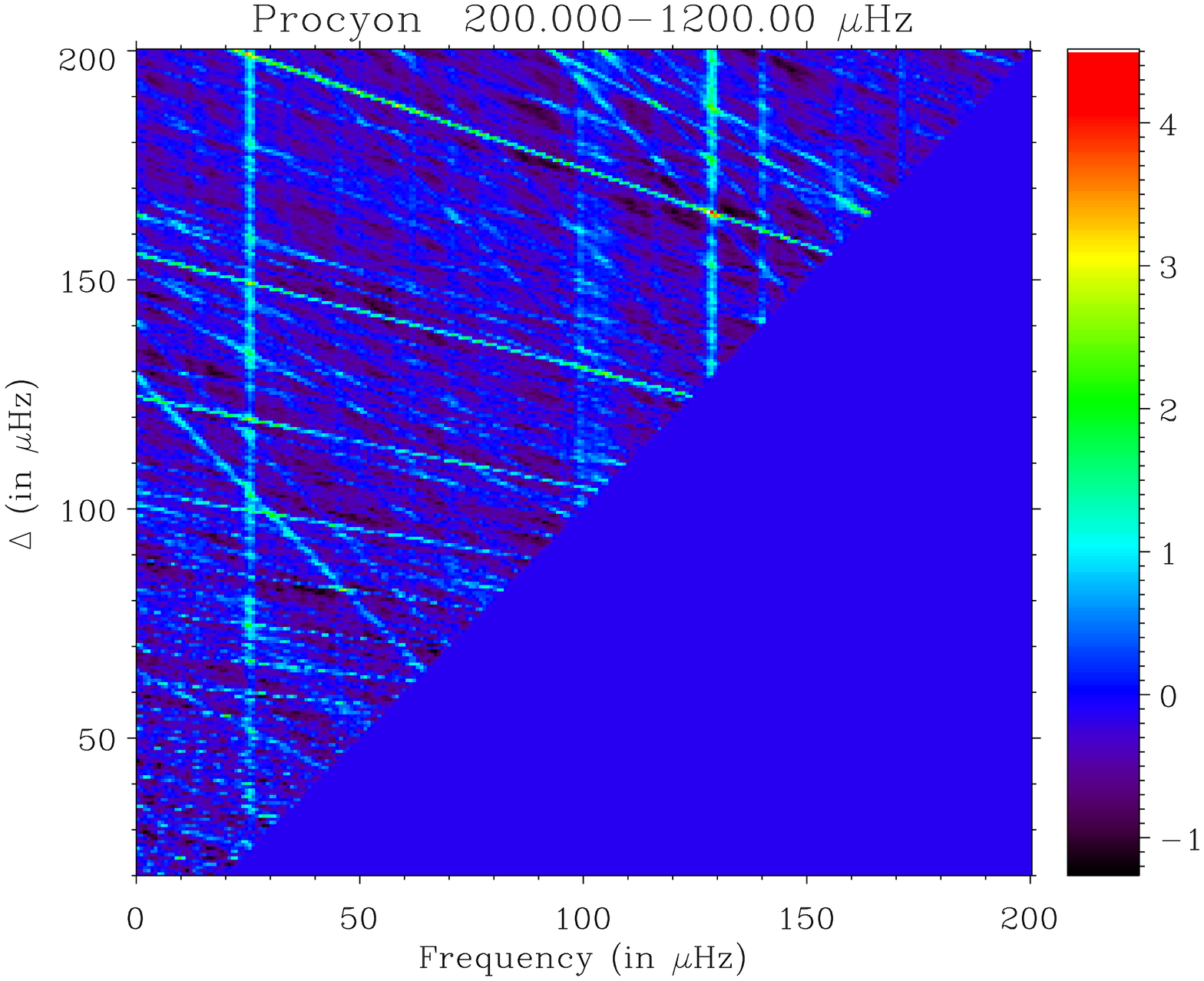}}
\caption{Left: Collapsed echelle diagramme for the Procyon window function. It is
dominated by a peak at $\simeq$\,165\,$\mu$Hz corresponding to the orbital
frequency, and other peaks following Equation~\ref{eq_spacing}. Right: Same diagramme
for Procyon data, with similar patterns}
\label{fig:procyonCED}
\end{figure*}

\begin{figure*}
\centering
\hbox{\includegraphics[height=8.5cm,angle=90]{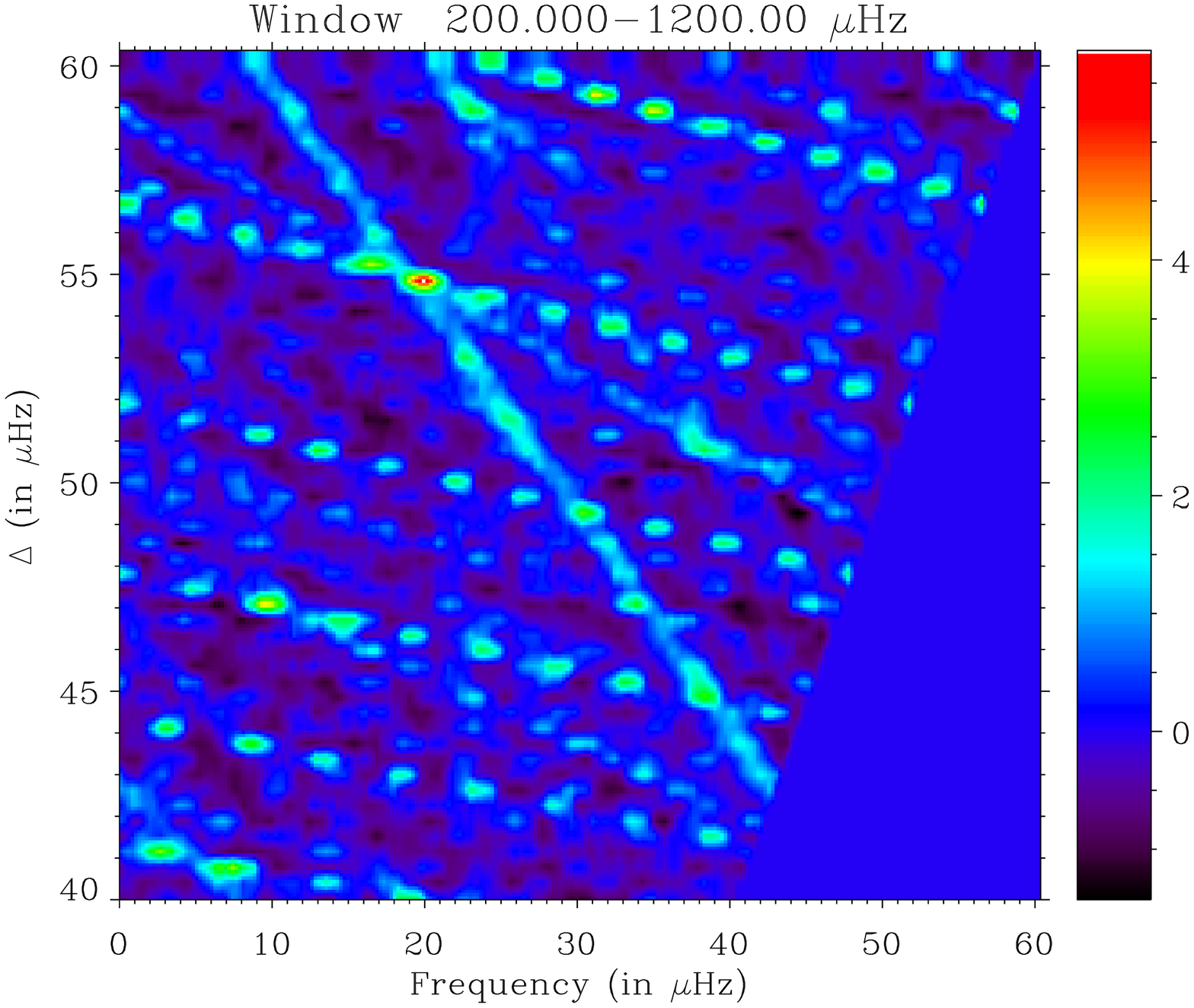}
\includegraphics[height=8.5cm,angle=90]{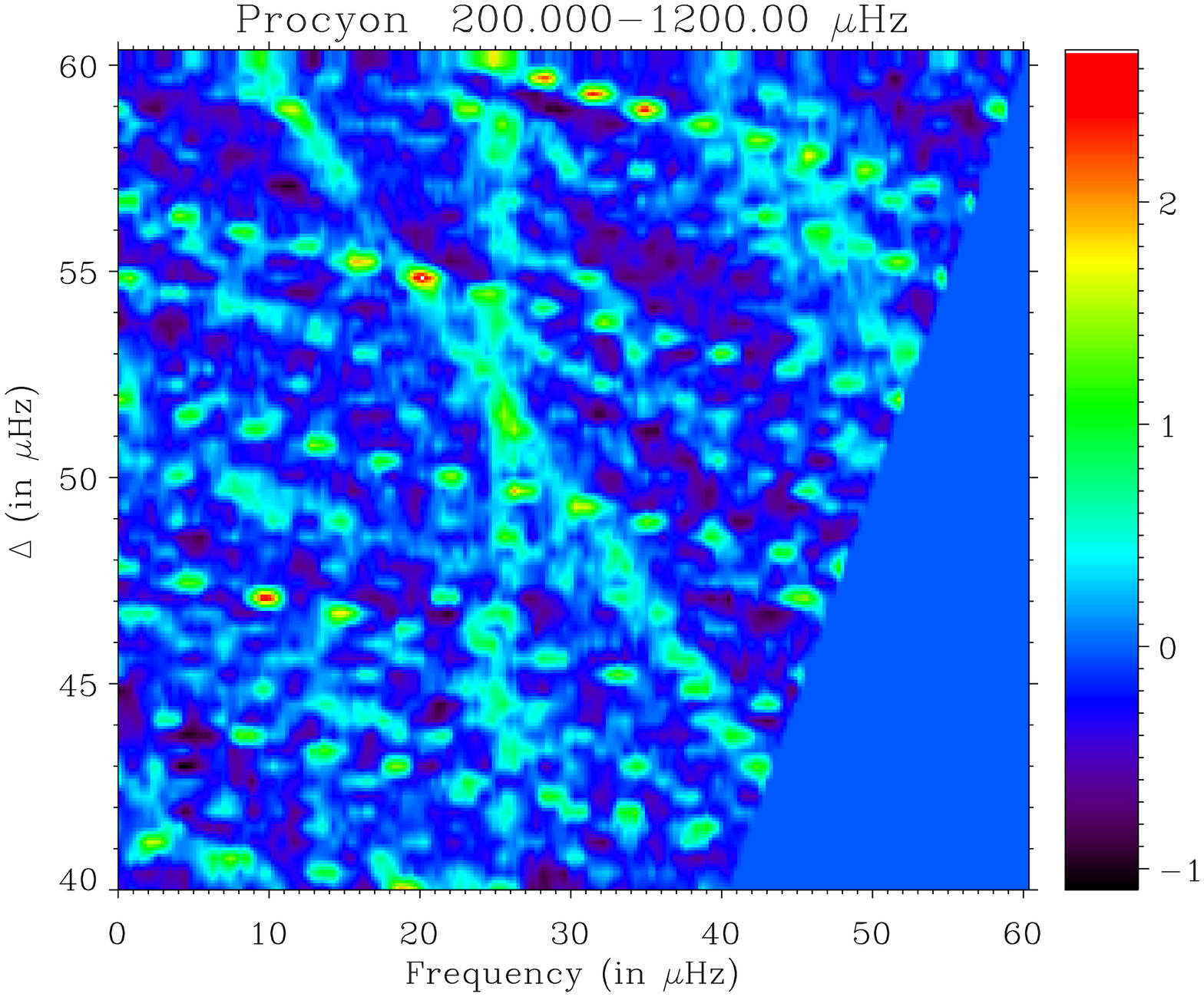}}
\caption{Respective enlargements of the Figure~\ref{fig:procyonCED}. The discontinuous
appearance is due do the discrete frequency sampling of the spectra.}
\label{fig:procyonzoom}
\end{figure*}

The same type of collapsed echelle diagrammes 
(with the same smoothing factor)			
were computed
for the MOST Procyon photometry. They have also been applied
to the window function of the MOST time series to assess its
influence on the analysis. The gaps in the data represent only
7.5 hours in a total observation of 32 days. This is illustrated by
the window function (see Fig.\ref{fig:win})
in which the presence of very short gaps which occur with some periodicity
(related to the satellite orbit), induces very weak peaks. Despite
their weakness, the regular spacing of these peaks in the Fourier spectrum
can interfere with the search for regularly-spaced p modes in this spectrum.
The left-hand panel of Figure~\ref{fig:procyonCED} shows the collapsed echelle
diagramme of the window function for the region of the
Fourier spectrum where stellar p~modes are expected. One can
see clear vertical and oblique structures: the vertical ones
are due to large peaks (whatever their origin) in the first
frequency interval used to build the diagramme, and the
oblique ones to similar large peaks, present in subsequent
frequency intervals. Clear peaks above the 99\% confidence
level appear at $\simeq$165, 110, 83, and 55 $\mu$Hz. The first
of these frequencies is the orbital frequency of the MOST
mission (\,1/101.4 minutes\,). The other three frequencies
correspond to the values of ${\Delta}'_0$ for ($n$~=~2, $m$~=~3),
($n$~=~1, $m$~=~2) and ($n$~=~1, $m$~=~3) respectively, taking 165
$\mu$Hz as ${\Delta}_0$ in Equation~\ref{eq_spacing}.\\
The same analysis was performed for the photometric time
series. The result, shown in the right-hand panel of Figure~\ref{fig:procyonCED},
shares the dominant pattern with the 2D collapsed echelle of
the window function, with ridges at the satellite orbital
frequency and for the frequency spacing defined by Equation~\ref{eq_spacing}.
Figure~\ref{fig:procyonzoom} shows enlarged views of the panels in
Figure~\ref{fig:procyonCED},
highlighting a small frequency range around 55 $\mu$Hz, which
is very close to 1/3 of the orbital frequency of MOST.  A
simple echelle diagram or comb analysis may yield some signal
at this frequency spacing which is not due to asymptotic
p~modes in Procyon but rather the intrinsic modulation of
stray light background in the photometry.

\section{Conclusion}

From the two approaches presented here, one can draw two
conclusions. First, the Procyon MOST light curve obtained in
2004 shows some changes in the spectral power content after about
seven days of observation (which were the first seven days of observation of the MOST
mission) that covers the frequency range
relevant to the p-mode search and thus requires special care
in this search. A counterpart to this change
in the CCD temperature telemetry indicates that it is not
intrinsic to Procyon.\\
Second, the echelle-diagramme
technique must be applied to Procyon MOST data with care. Despite an
excellent observation window, 
the orbital artifacts in the data introduce apparent spacings
close to the p-mode spacing expected for Procyon even though the gaps
are very short and their signature in the window spectrum is very weak. However,
they are unexpectedly not negligible when searching for regularly--spaced
peaks in the Fourier spectrum, reinforcing the need for a careful
analysis when looking for regular spacings in the signal spectrum even in the
case of a very good observation window.\\
In conclusion, we do not find any evidence for a stellar oscillation signal
in the MOST photometry, and the issue of the nature of
luminosity oscillations in Procyon remains open.

\begin{acknowledgements}
T.A. wishes to thank M.\,Marti\'c and S. Korzennik for helpful discussions.
J.M.M. acknowledges funding from the Natural
Sciences \& Engineering Research Council (NSERC) Canada. R.K.
is supported by the Canadian Space Agency (CSA).
\end{acknowledgements}

\bibliographystyle{aa}
\bibliography{procyonv4}

\end{document}